\documentclass[preprint,5p,number,twocolumn]{elsarticle}
\usepackage{lineno}

 \usepackage{graphics}
 \usepackage{graphicx}

\usepackage{amssymb}
\usepackage{amsthm}
\usepackage[FIGBOTCAP]{subfigure}
\usepackage{lineno}
\usepackage{amsmath,amssymb} 
\usepackage[pdftex]{hyperref}

\journal{Elsevier}

\begin{document}

\begin{frontmatter}



 \cortext[cor1]{Corresponding author}

\title{\Large\bf A beam monitor using silicon pixel sensors for hadron therapy}

\author{Zhen Wang}
\ead{zwang@mails.ccnu.edu.cn}
\author[]{Shuguang Zou}
\author[]{Yan Fan}
\author[]{Jun Liu}
\author{Xiangming Sun$^{*}$}
\ead{sphy2007@126.com}
\author[]{Dong Wang}
\author[]{Huili Kang}
\author[]{Daming Sun}
\author[]{Ping Yang}
\author[]{Hua Pei}
\author[]{Guangming Huang}
\author[]{Nu Xu}
\author[]{Chaosong Gao}
\author[]{Le Xiao}
\address[ccnu]{PLAC, Key Laboratory of Quark $\&$ Lepton Physics (MOE), Central China Normal University, Wuhan, Hubei 430079, PR China}

\begin{abstract}
We report the design and test results of a beam monitor developed for online monitoring in hadron therapy. The beam monitor uses eight silicon pixel sensors, \textit{Topmetal-${II}^-$}, as the anode array. \textit{Topmetal-${II}^-$} is a charge sensor designed in a CMOS 0.35 $\mu$m technology. Each \textit{Topmetal-${II}^-$} sensor has $72\times72$ pixels and the pixel size is $83\times83$ $\mu$m$^2$. In our design, the beam passes through the beam monitor without hitting the electrodes, making the beam monitor especially suitable for monitoring heavy ion beams. This design also reduces radiation damage to the beam monitor itself. The beam monitor is tested with a carbon ion beam at the Heavy Ion Research Facility in Lanzhou (HIRFL). Results indicate that the beam monitor can measure position, incidence angle and intensity of the beam with a position resolution better than 20 $\mu$m, angular resolution about 0.5$^\circ$ and intensity statistical accuracy better than 2$\%$.

\end{abstract}


\end{frontmatter}

\section{Introduction}\label{Introduction}

Cancer treatment with hadron beams has seen rapid developments in clinical use over the past 10-15 years\cite{TumorTherapy}. Today, about 70 proton and carbon-ion therapy centers are in operation all over the world\cite{facility}. When an ion travels through matter, most of its energy is deposited very close to the end of the trajectory. This is called the Bragg-peak mechanism. By modulating the ion kinetic energy, a local high dose can thus be delivered even to deep-seated the tumors. The dose delivered to the surrounding healthy tissue is typically reduced compared to X-ray radiation therapy and can be minimized further by varying the incident direction of the ion beam.

A standard hadron therapy facility consists of ion sources, an accelerator to bring the ions to hundreds of $MeV/u$, and beam delivering systems. In the active spot-scanning delivery system, scanning magnets deflect the pencil beam to the required transverse position and the accelerator modulates the beam energy to deliver the Bragg peak of the beam to the required longitudinal position\cite{PSI}.

To ensure accurate delivery of a prescribed dose to the tumor, a beam monitor system is required to measure the beam's intensity, position, and profile in real time. The beam monitor must also have low material budget to reduce the disturbances to the beam. 

Currently, parallel-plate ionization chambers with one large electrode or electrodes segmented in strips\cite{PSI,HIRFL_IonChamber,INFN,CNAO,TOP-IMPLART} or pixels\cite{CNAO} are commonly used at hadron therapy facilities to monitor the beam online. The ionization chambers can measure the beam with a sub-millimeter position resolution and with an intensity accuracy of about 1$\%$, which are general requirements of hadron therapy\cite{PSI,CNAO}. Moreover, a new device, a Monolithic Active Pixel Sensor (MAPS), is under development for the monitoring of the ion beam\cite{maps}.

We are developing an online beam monitor that is based on the TPC (time projection chamber) detection principle\cite{TPC}. Eight silicon pixels sensors, \textit{Topmetal-${II}^-$}\cite{topmetal-II-}, are used as an anode array. Here, we demonstrate that this beam monitor measures the position of the beam with a better than 20 $\mu$m resolution and the angle of the beam with about 0.5$^\circ$ resolution. These characteristics could, in the future, improve the accuracy of hadron therapy.

In part 2 we describe the silicon pixel charge sensor \textit{Topmetal-${II}^-$} and the design of the beam monitor that uses this sensor. We show measurement results at HIRFL in part 3 and a comparison with ionization chambers in part 4.

\section{Design of the beam monitor}

\textit{Topmetal-${II}^-$} is a charge sensor, fabricated using a CMOS 0.35 $\mu$m technology\cite{topmetal-II-}. The sensor has $72\times72 = 5184$ pixels. The pixel size is $83\times83$ $\mu$m$^2$. This leads to a charge sensitive region of $6\times6$ mm$^2$.

Each pixel has a $25\times25$ $\mu$m$^2$ exposed top layer metal pad that collects the charge (hence the name Topmetal), a charge sensitive amplifier (CSA), an analog readout channel and a digital readout channel. The CSA converts the charge into an analog voltage signal. As we only use the the analog readout of the senor in the monitor, we will not discuss the digital readout in this paper. In the analog readout, the voltage signal from the CSA goes through two pixel level source-follower stages and then is routed through a chip level analog buffer to be read out. The analog Equivalent Noise Charge (ENC) of the sensor is measured to be below 15 e-. The sensor response to injected charges is linear\cite{tm2-_lowtemp}.
    
\begin{figure} [htbp]
		\centering
		\includegraphics[width=0.95\columnwidth]{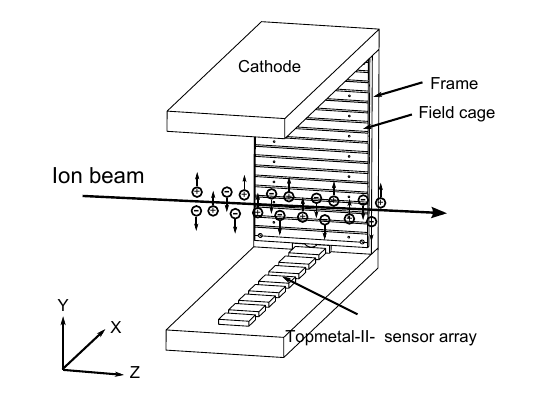}
		\centering
		\caption{Schematic view of the monitor(the foreground half of the field cage and frame are hidden). The eight \textit{Topmetal-${II}^-$} sensors are lined as the anode array. The field cage is used to make the electric field perpendicular to electrodes. The anode array plane is defined as the x-z dimension. The beam passing through the monitor is projected on the x-z dimension.}	
		\label{monitor}		
	\end{figure}

We employ the TPC (time projection chamber) detection principle\cite{TPC} in our beam monitor design, shown in Fig.\ref{monitor}. The beam monitor locates at the end of the beam line, in ambient air and at room temperature. When the hadron beam passes through the electrodes, electrons and ionized air molecules are generated and start to drift under the influence of the applied electric field. The electrons are quickly captured by neutral oxygen or water molecules in the air, forming negative ions\cite{negativeIon}, which are then detected on the \textit{Topmetal-${II}^-$} sensor acting as the anode of the TPC. The slow drift velocity and heavy mass of these ions (compared with electrons used in ionization chambers) contribute to the improvements in measurement resolution. The slow drift velocity also better matches the shaping time in \textit{Topmetal-${II}^-$}. Alternatively, with opposite polarity of the electric field, the positive ions could be directly detected. Note that due to the diffusion of the ions, the distribution of the negative ions should be somewhat broader than the initial beam, but this effect can be corrected.
 
In the final design, it is planned to place two monitors in the beam with the drift fields orthogonal to each other. Each monitor provides a 2-dimensional projection of the beam on the anode plane. The beam profile and the incidence angle can then be reconstructed from the two orthogonal projections.

Shown in Fig.\ref{monitor}, eight \textit{Topmetal-${II}^-$} sensors are lined up to form the anode of the TPC (the monitor). The sensor area including the readout circuits is $8\times9$ mm$^2$, the sensitive region is $6\times6$ mm$^2$. This leads to a dead area of 5.4 mm between the sensors. The anode is set to the ground, while the cathode, which is 6 cm from the anode, is at -2000 V potential. The field cage is used to make the electric field perpendicular to the electrodes\cite{fieldcage}. The analog readout of the sensor uses a rolling shutter method with a 1.5625 MHz clock. The eight sensors are read out in parallel, however, in this study we show data from only one sensor, because scanning of the ion beam was not possible at the time of the experiment. The readout time for one frame of the whole monitor is about 3.3 ms.

\section{Measurements at HIRFL}
We tested the monitor in the former superficially-placed tumor therapy terminal at the Heavy Ion Research Facility in Lanzhou (HIRFL) China\cite{HIRFL_BeamDeliver}. The scanning magnets in the beam delivery system are shut down and the beam delivery system delivers a 80.55 $MeV/u$ carbon ion beam. The beam is time-continuous. The beam passes through a 30 mm long collimator and then through the monitor. The collimator has two different inner diameters of 2 mm and 3 mm to choose. Because in the monitor there are gaps between the sensors, and the ion beam can not be moved, the position of the monitor is adjusted manually to ensure that the ion beam is well centered on one sensor.

\subsection{Two-Dimensional Projection of the Beam} 
Every 3.3 ms the monitor can provide a two-dimensional projection of the beam, thus measuring its position, incidence angle and intensity. 
 
\begin{figure} [htbp]
	\centering
	\includegraphics[width=0.95\columnwidth]{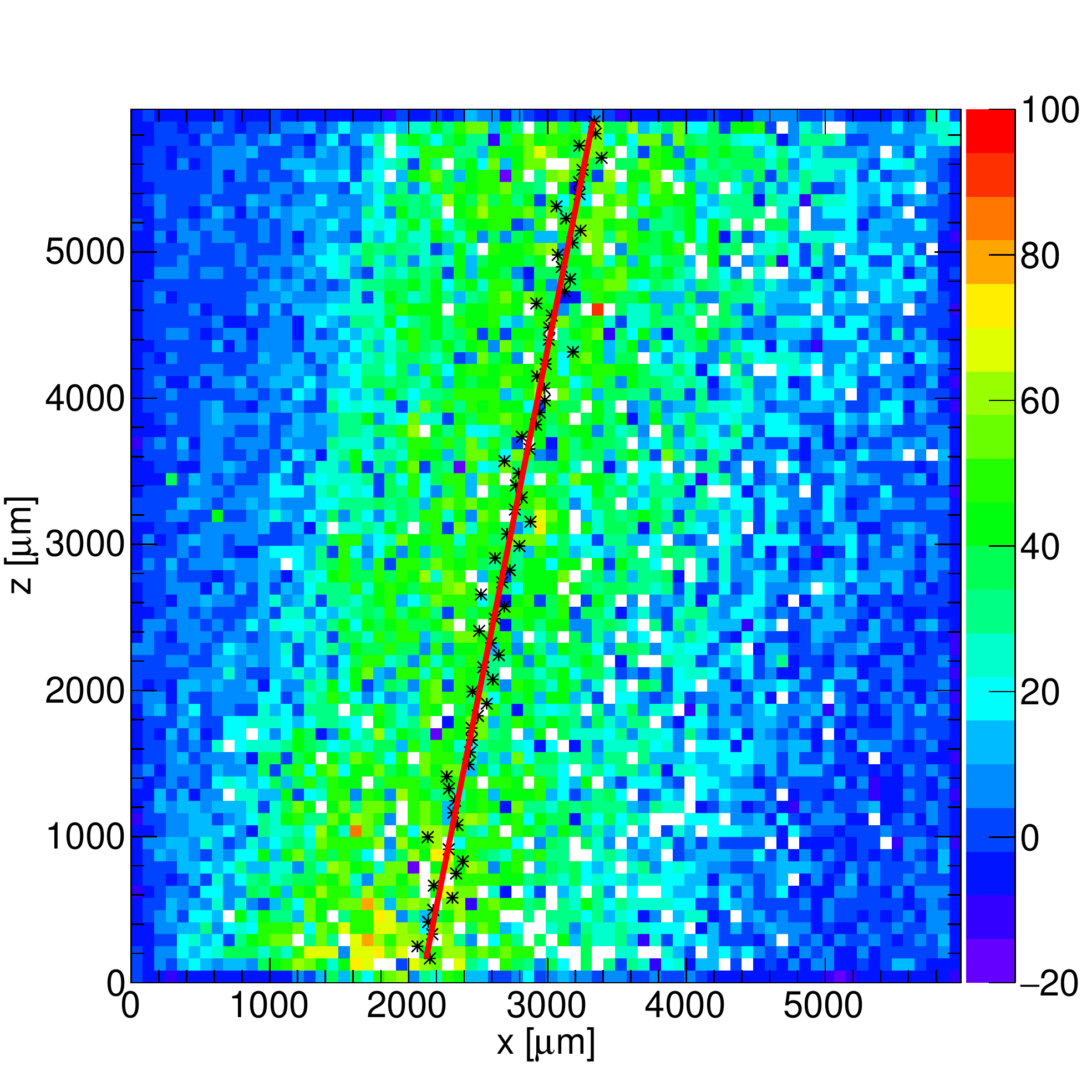}
	\centering
	\caption{Two-dimensional projection of the 80.55 $MeV/u$ carbon ion beam passing through the collimator with an inner diameter of 2 mm. The data are from a single 3.3 ms readout cycle. The black point is the center of gravity of each row. The red line shows the linear fit of the black points. The angle between the red line and the positive z-axis is defined as the incidence angle of the beam. The incidence angle is about 9$^\circ$. The sum of the response of all pixels is defined as the intensity of the beam. The two-dimensional projection of the beam is not smooth, which results from the nonuniformity of pixels. The white bins are bad pixels with a too large or too small decay time constant.}	
	\label{beam_Angle9}		
\end{figure}

In the test, the carbon ion beam passes through the collimator with a 2 mm inner diameter and then the monitor. The two-dimensional projection of the beam on the monitor is shown in Fig.\ref{beam_Angle9}. The data is from a single 3.3 ms readout slice. The center of gravity of each row of the two-dimensional projection is calculated by $$X_{cog} = \frac{\sum\limits_{i=1}^{72}{x_is_i}}{\sum\limits_{i=1}^{72}{s_i}},$$ where $s_i$ is the response of the pixel in the $i_{th}$ column and $x_i$ is the coordinate of the pixel in the $i_{th}$ column.  The centers of gravity from 72 rows are fitted linearly, as the red line in Fig.\ref{beam_Angle9} shows. Here, we assume the beam is a straight line and the fitted red line represents the beam. The fitted red line is defined as the center position of the beam in each row. The angle between the red line and the positive z-axis is defined as the incidence angle of the beam. For the given position of the monitor, the incidence angle is about 9$^\circ$. The sum of the response of all pixels is defined as the intensity of the beam.  

Note that in Fig.\ref{beam_Angle9} some white bins exist. The corresponding pixels are considered bad pixels. The CMOS technology is unable to make the internal circuits of all pixels identical, leading to different decay time constants. If the decay time constant of a pixel is too small (in this case $<$3.3 ms), the response of the pixel would disappear rapidly. Such pixels are called cold pixels. If the decay time constant of a pixel is too large, the pixel would be saturated when continuous charges are collected. Such pixels are called hot pixels. However, the classification of hot pixels not only depends on the decay time constant, but also on the collected charge. Both, cold and hot pixels have been removed in Fig.\ref{beam_Angle9} for the calculation of the position ($X_{cog}$) and the beam intensity. Moreover, the different decay time constants of the pixels lead to a nonuniformity\cite{tm2-_lowtemp}, meaning that the responses of the pixels are different while the pixels collect the same charges. This causes the two-dimensional projection of the beam shown in Fig.\ref{beam_Angle9} to be not completely smooth.

\begin{figure} [htbp]
	\centering
	\includegraphics[width=0.95\columnwidth]{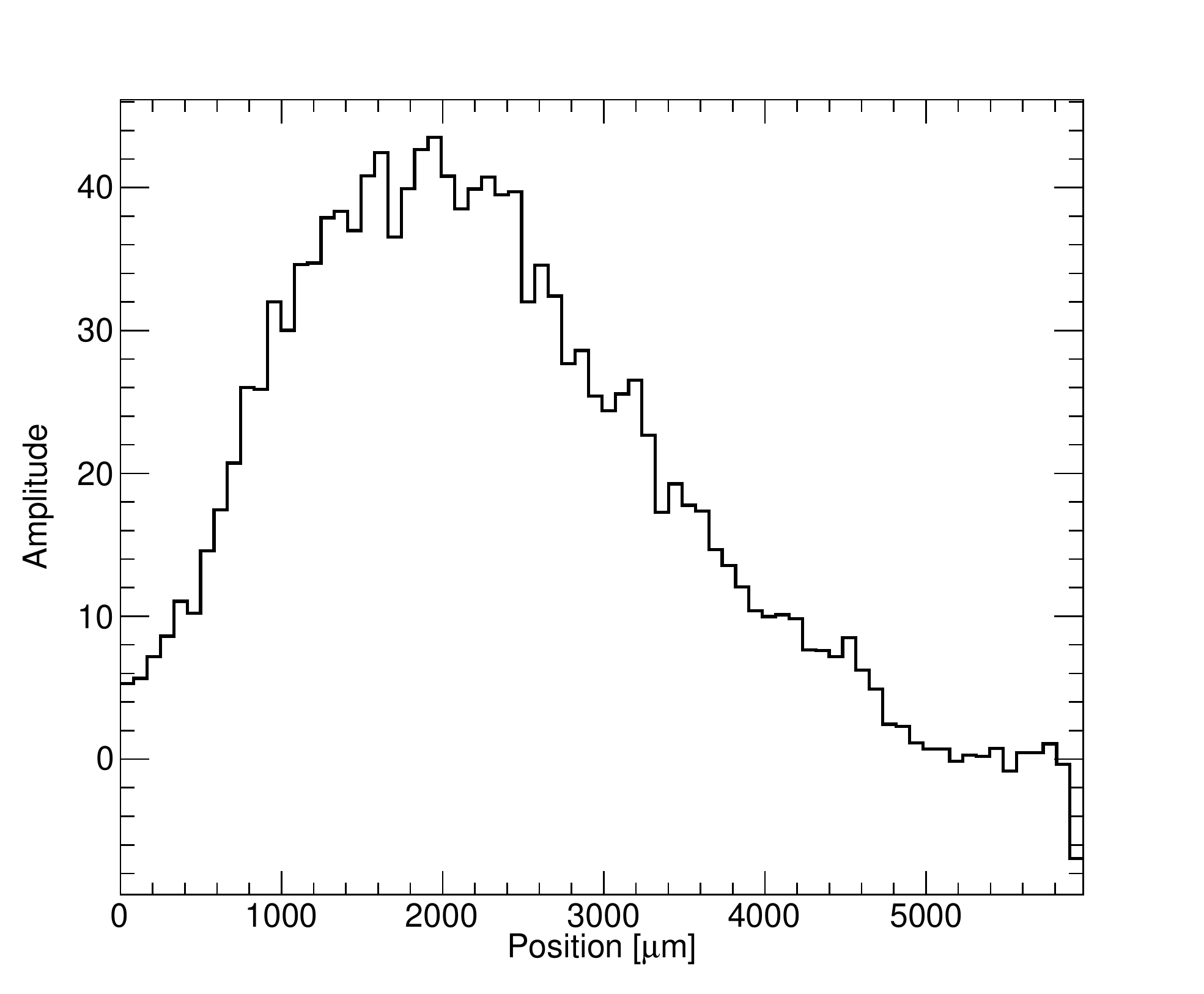}
	\centering
	\caption{One-dimensional projection of Fig.\ref{beam_Angle9} onto the axis perpendicular to the fitted red line.}	
	\label{beamProfile}		
\end{figure}

In Fig.\ref{beamProfile} we show the one-dimensional projection of Fig.\ref{beam_Angle9} onto the axis perpendicular to the fitted red line, corresponding to the beam profile in x direction. 

\subsection{Performance of the Monitor}	
The monitor is designed to measure the fluctuation of the beam position, beam angle and beam intensity. The performance is studied with a continuous data acquisition. In this test the beam passes through the collimator with a 3 mm inner diameter and then the monitor in the direction almost parallel to z. The monitor samples 8089 frames during 27 s. These data are used to show the fluctuation of the beam as a function of time and analyze the position resolution, angular resolution, intensity accuracy of the monitor.  

\begin{figure} [htbp]
 		\centering
        \subfigure[]{
            \includegraphics[width=0.95\columnwidth]{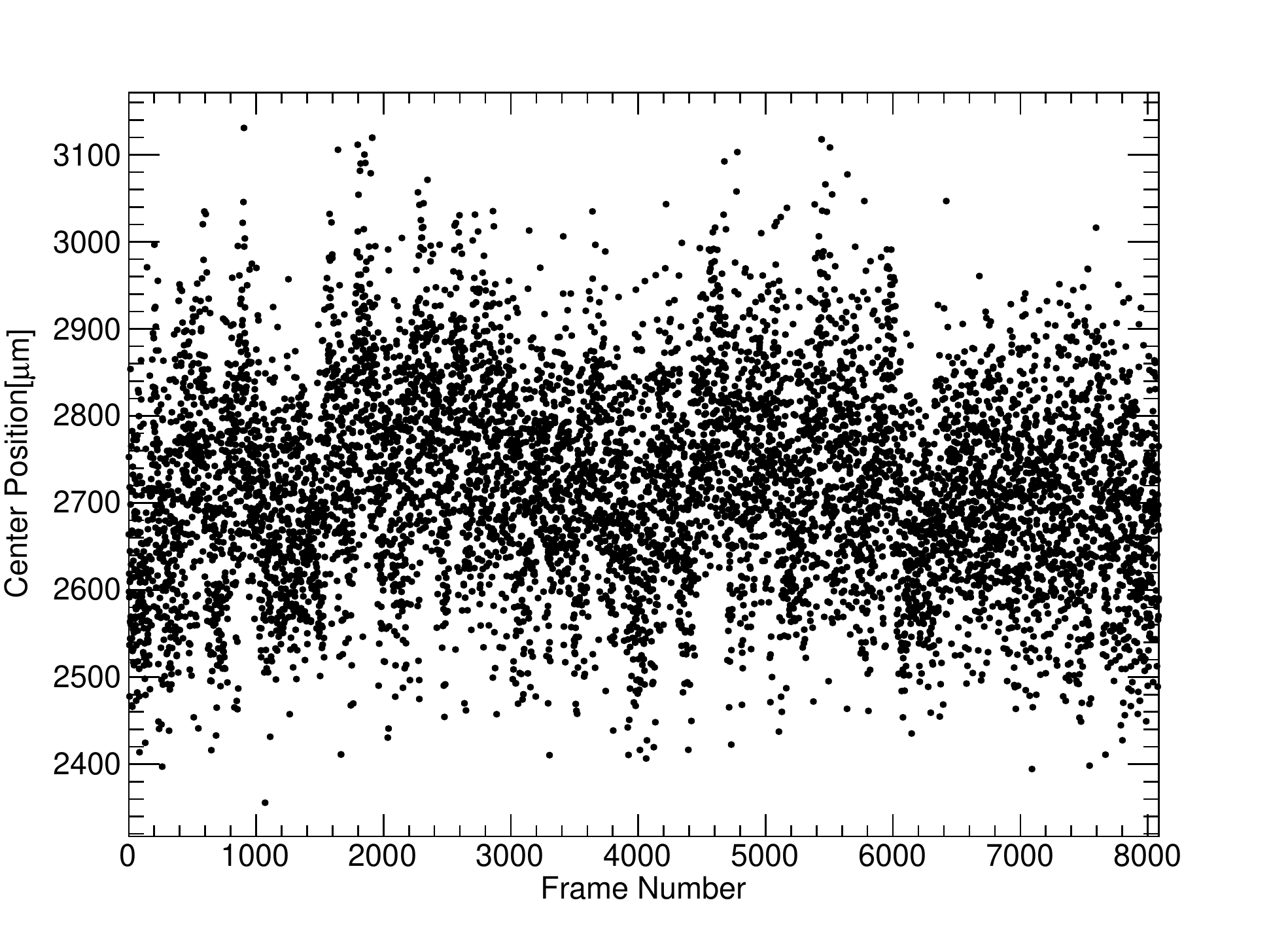}        
        } 
        \subfigure[]{
            \includegraphics[width=0.95\columnwidth]{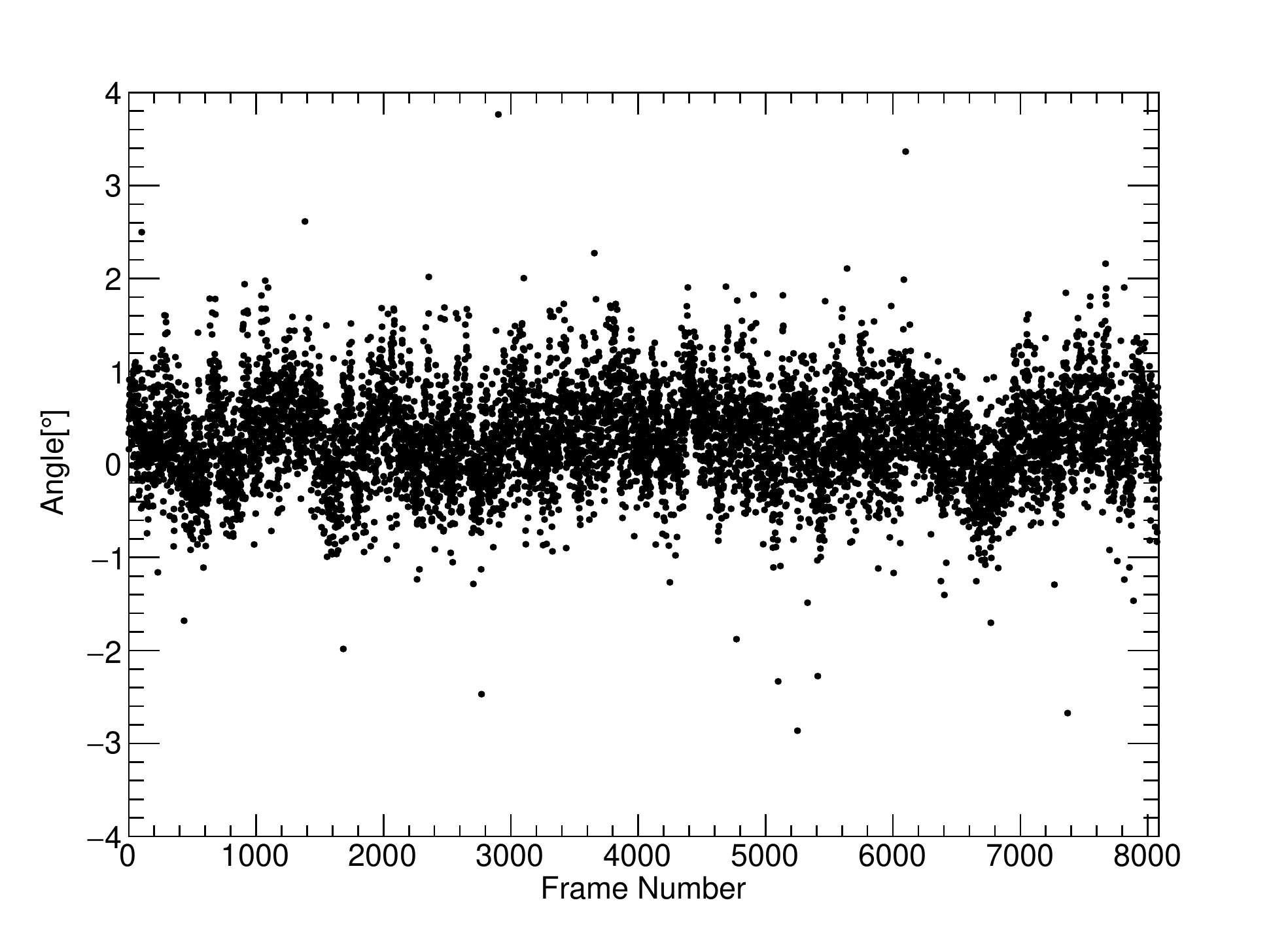}        
        } 
        \subfigure[]{
            \includegraphics[width=0.95\columnwidth]{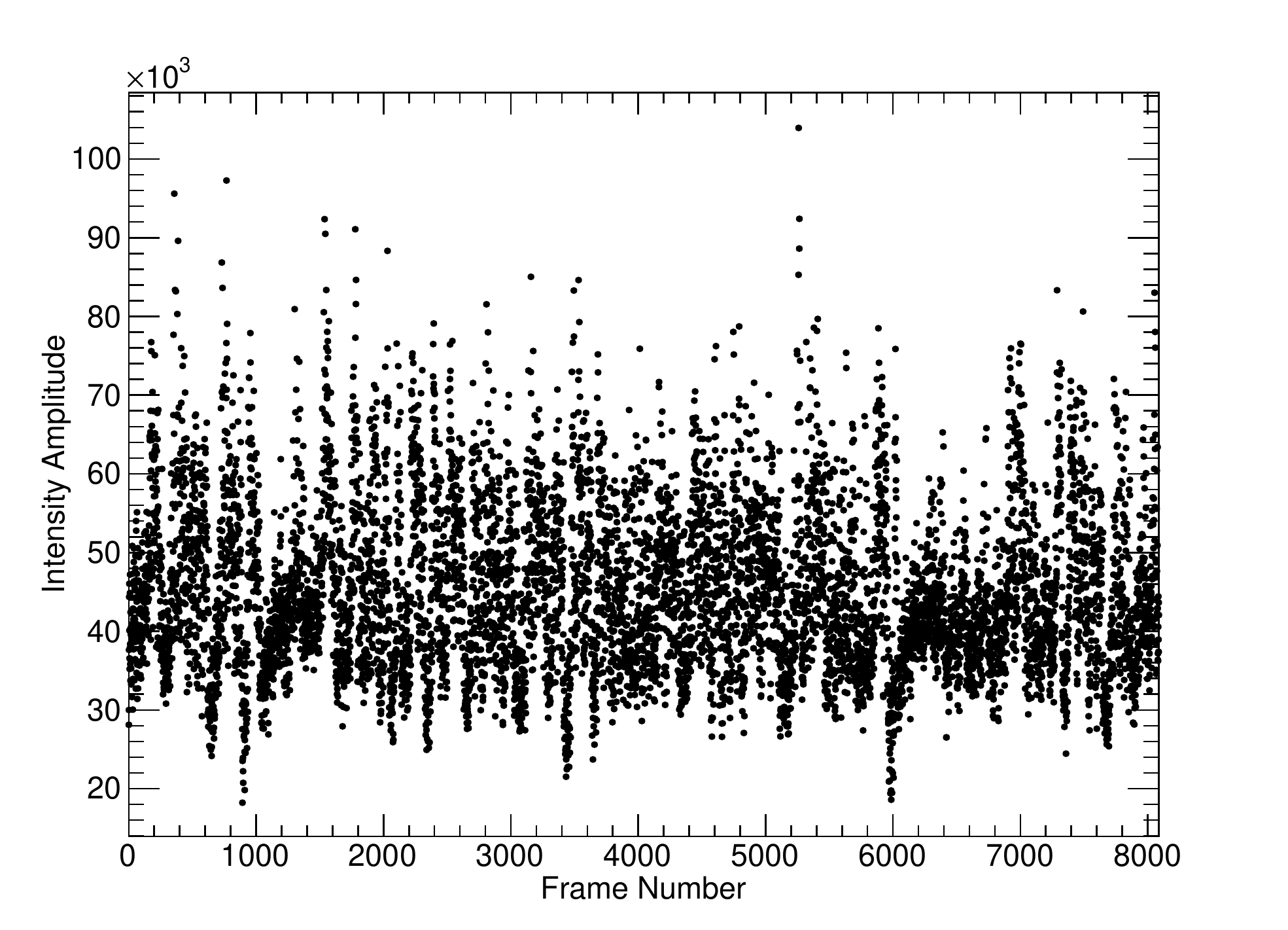}        
        }                 
 		\centering
 		\caption{Center position(a), angle(b) and intensity(c) of the beam as a function of time. One frame represents 3.3 ms. The beam passes through the collimator with an inner diameter of 3 mm and then the monitor in the direction almost parallel to z. The incidence angle values of the beam are around 0$^\circ$.  }	
 		\label{beamVStime}		
 	\end{figure}

In Fig.\ref{beamVStime} we show the center position, angle and intensity of the beam as a function of the frame number. One frame represents 3.3 ms.

\begin{figure} [htbp]
 		\centering
 		\includegraphics[width=0.95\columnwidth]{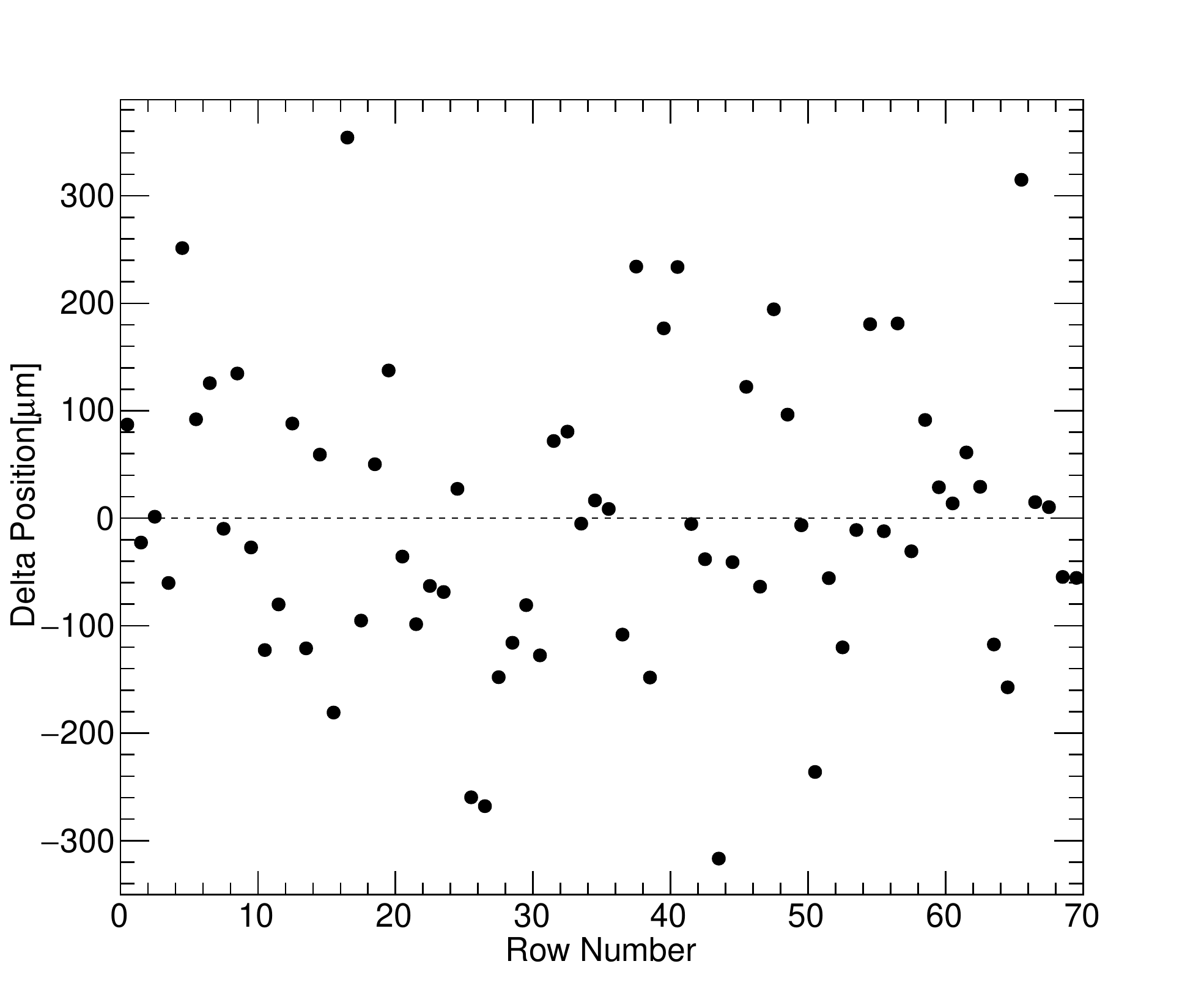}
 		\centering
 		\caption{Delta position of each row in one frame. Delta position is the distance vector between the center of gravity (black point) of each row and the fitted red line. }	
 		\label{deltaPosition}		
 	\end{figure}

The beam monitor position resolution is a paramount parameter to fulfill the accurate beam delivering. In the test we don't have a calibrated position detector to measure the beam position independently, and the position fluctuates strongly from frame to frame, as is shown in Fig.\ref{beamVStime}. Therefore, we adapt the following method to calculate the position resolution. We assume the beam is a straight line, has a stable position in one frame, and the fitted red line represents the beam. The distance vector between the center of gravity (black point) of each row and the fitted red line is defined as the delta position. For one frame, the delta position of each row is shown in Fig.\ref{deltaPosition}. The standard error of the delta position is defined as the position resolution. 

\begin{figure} [htbp]
 		\centering
 		\includegraphics[width=0.95\columnwidth]{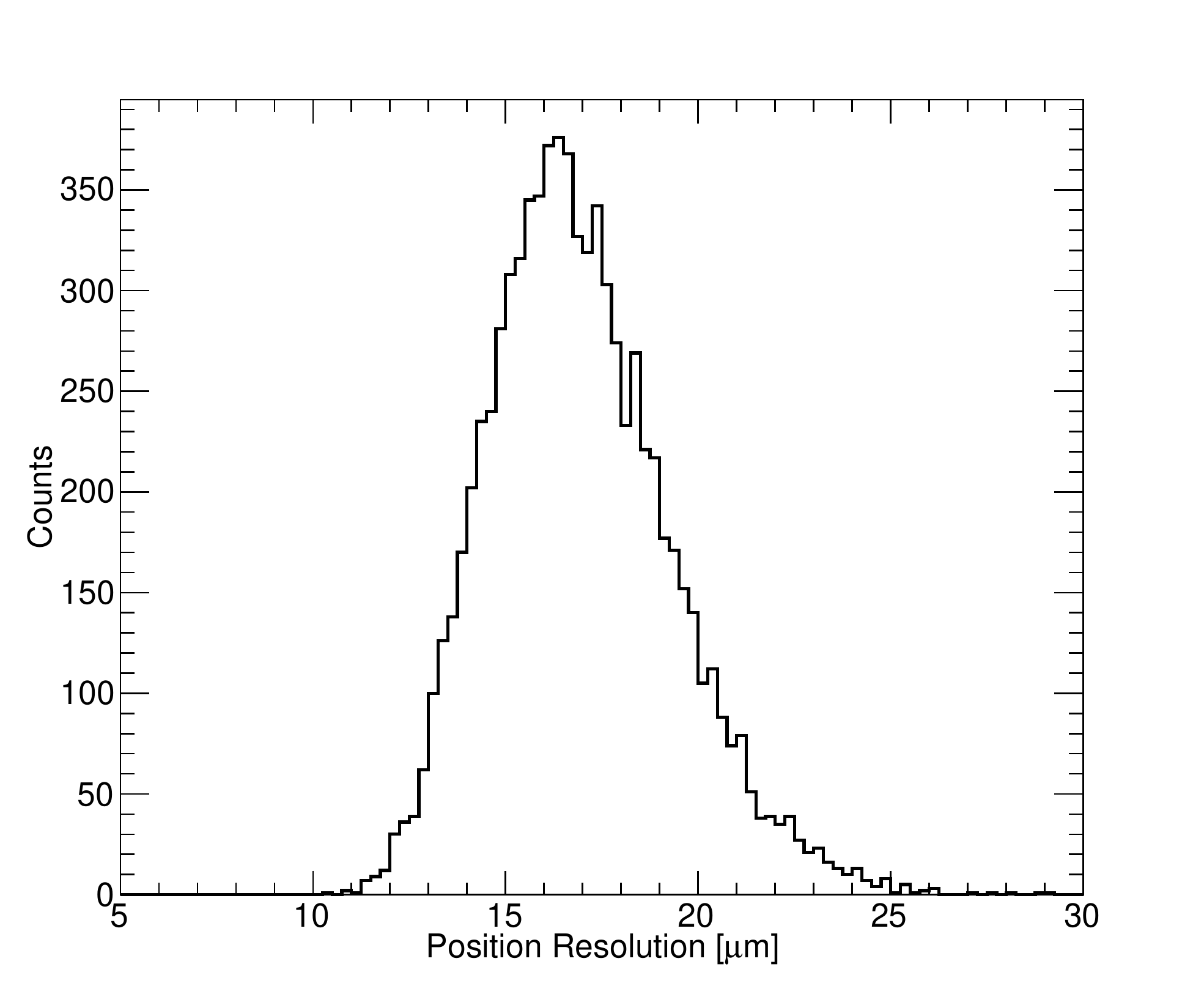}
 		\centering
 		\caption{Distribution of the position resolution of the 8089 frames. Most probable value of the position resolution is 17 $\mu$m.     }	
 		\label{positionResolution_Angle0}		
 	\end{figure}

In the monitor, each frame can give the beam position resolution.  The distribution of the position resolution for the 8089 frames is shown in Fig.\ref{positionResolution_Angle0}. The most probable value of the position resolution is 17 $\mu$m.
 	
\begin{figure} [htbp]
  		\centering
  		\includegraphics[width=0.95\columnwidth]{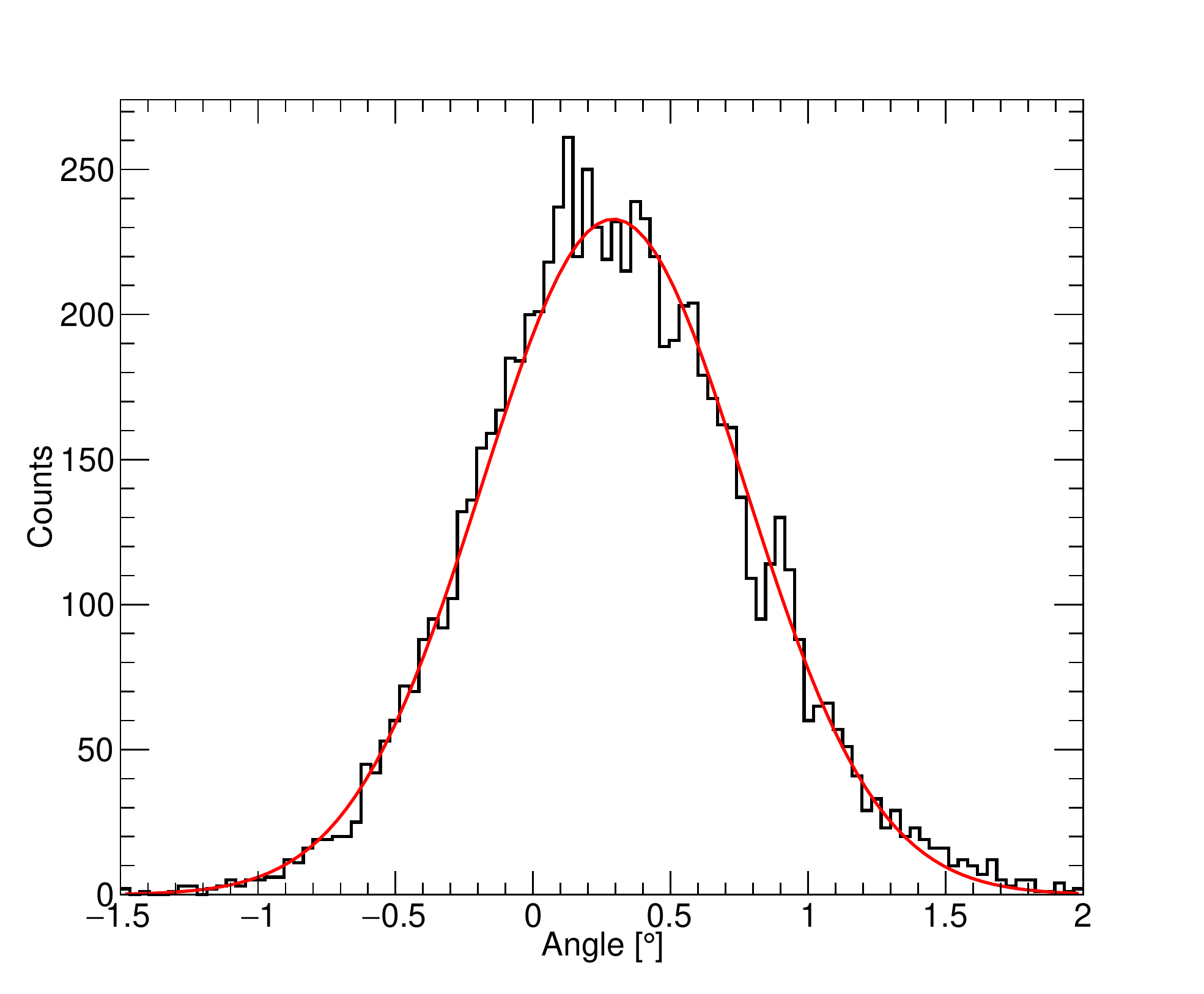}
  		\centering
  		\caption{Distribution of the beam angle of the 8089 frames. The angular resolution is about 0.5$^\circ$.  The shape is fitted with a Gaussian shape and it is shown here as red curve for eye-directing purpose.   }	
  		\label{angleResolution_Angle0}		
  	\end{figure}

The angle between the red line and the positive z-axis is defined as the incidence angle of the beam. Each frame can give a beam angle. The distribution of the beam angles for the 8089 frames is shown in Fig. \ref{angleResolution_Angle0}. The RMS of the distribution of the beam angle is defined as the angular resolution. The angular resolution is about 0.5$^\circ$.
  	
\begin{figure} [htbp]
 		\centering
 		\includegraphics[width=0.95\columnwidth]{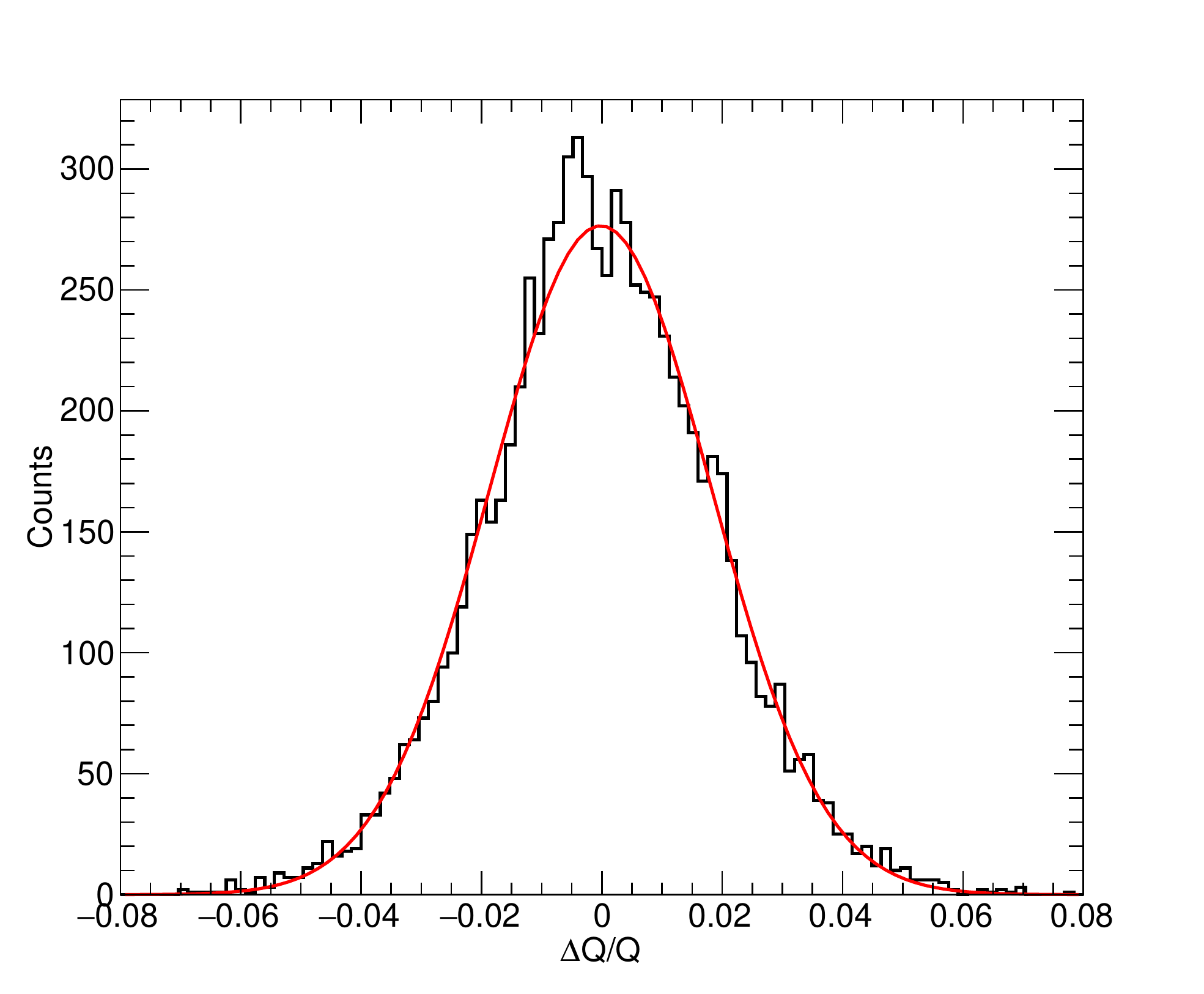}
 		\centering
 		\caption{Distribution of $\Delta Q/Q$  of the 8089 frames. $Q_{up}$ is the summation of the charges of all the pixels in the upper 36 rows and $Q_{down}$ is the summation of the charges of all the pixels in the lower 36 rows in one frame. $\Delta Q=Q_{up}-Q_{down}$, $Q=(Q_{up}+Q_{down})/2 $. The RMS of the distribution is 2$\%$ which is the beam intensity statistical accuracy. The shape is fitted with a Gaussian shape and it is shown here as red curve for eye-directing purpose.}	
 		\label{intensityResolution_Angle0}		
 	\end{figure}

The dose deposition to the tumor largely depends on the beam intensity and the beam from the accelerator is time-varying. We adapt the following method to calculate the intensity accuracy of the monitor. The \textit{Topmetal-${II}^-$} contains 72 rows, which are divided into two parts, each part of 36 rows. We assume in one frame the beam doesn't vary and the beam intensity in the upper part and lower part is the same. That means the beam is measured twice repeatedly. In one frame, the charges of all the pixels in the upper part are summed as $Q_{up}$ and that in the lower part are summed as $Q_{down}$. $\Delta Q=Q_{up}-Q_{down}$, $Q=(Q_{up}+Q_{down})/2 $. The distribution of $\Delta Q/Q$ of the 8089 frames is shown in Fig.\ref{intensityResolution_Angle0}.
The mean of the distribution of the $\Delta Q/Q$ approximates zero.  The RMS of the distribution of the  $\Delta Q/Q$, defined as the beam intensity statistical accuracy, is 2$\%$. 
\section{Comparison with Ionization Chamber}
The monitor based on silicon pixel sensors can provide better than 20 $\mu$m position resolution and measure the angle of the beam with about 0.5$^\circ$ resolution. For parallel-plate ionization chambers, the mm-scale width of strips typically restricts the position resolution of that dimension to $>$0.1 mm.   

\begin{figure} [htbp]
 		\centering
 		\includegraphics[width=0.95\columnwidth]{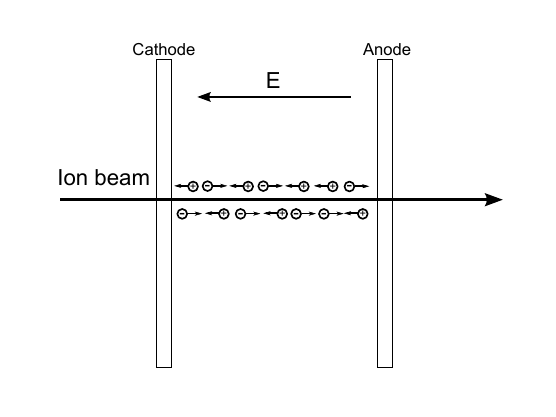}
 		\centering
 		\caption{Schematic view of the parallel-plate ionization chamber. The beam passes through the electrodes. The ions and electrons are produced along the direction of beam and then move under the electric field. The ions moving along the beam have a significant probability of recombining with the electrons. This signal lost due to recombination will result in inaccurate measurement of beam intensity.}	
 		\label{IonChamber-paper}		
 	\end{figure}

Moreover, in the parallel-plate ionization chamber, the ions and electrons are produced along the direction of the beam and then move under the electric field, in parallel with the direction of the beam, as shown in Fig.\ref{IonChamber-paper}. The ions moving along the beam have a significant probability of recombining with the electrons then. With the pencil beam size smaller and intensity higher, the density of ions and electrons produced along the beam also becomes larger, which results in a higher recombination rate. This signal lost due to recombination will result in inaccurate measurement of beam intensity\cite{negativeIon,recombination}. For the monitor we are proposing, the ions and electrons produced drift under an electric field which is perpendicular to the direction of the beam, see Fig.\ref{monitor}. Thus, the recombination rate of the ions and  electrons should be less compared to ionization chambers. Therefore, the monitor proposed in this paper contributes to reduce the recombination rate for the sharp pencil beam. 
   
\section{Conclusion and Outlook}

The monitor using the silicon pixel sensors \textit{Topmetal-${II}^-$} has been developed for online monitoring in hadron therapy. Different from the ionization chamber, the monitor is based on the TPC detection principle. The monitor provides better than 20 $\mu$m position resolution which is higher than the current ionization chambers. In principle, the monitor contributes to reduce the recombination rate for the sharp pencil beam. It provides about 0.5$^\circ$ angular resolution.

There is an on-going updated design of this monitor, which includes 2 sets of lined sensor arrays in perpendicular angle. This will provide the 2-dimensional profile of beams injected and passing over both arrays with good resolution in both x-z and y-z dimension. Furthermore, this new design eliminates possible effects of dead area between neighboring sensors, since the array in each dimension contains two lines of eight sensors aligned staggered (32 sensors in total), such that whenever the beam passes over space between sensors in one line, it will be caught by one sensor in the neighboring line.

In the active spot-scanning delivery system, the capability of instantaneous online monitoring of the beam energy is very important. The deflection angles of the ions of different energies are different in the magnet field. We can measure the beam position before and after the magnet field. By comparing the difference of the beam positions before and after the magnet field, we should get the energy profile of the beam. The silicon pixel sensor \textit{Topmetal-${II}^-$} with a pixel size of 83 $\mu$m is suitable for position sensitive detectors. 
   
\section*{Acknowledgments}
	This work is supported, in part, by the Thousand Talents Program at Central China Normal University and  by the National Natural Science Foundation of China under Grant No.11375073, No.11305072, No.1232206 and No.11220101005.



\begin{thebibliography}{00}












\bibitem{TumorTherapy}
G. Kraft, Tumor therapy with heavy charged particles, Prog. Part Nucl. Phys. 45, 473 (2000); http://dx.doi.org/10.1016/S0146-6410(00)00112-5 
\bibitem{facility}
http://www.ptcog.ch/index.php/facilities-in-operation
\bibitem{PSI}
O. Actis, D. Meer and S. König, Precise on-line position measurement for particle therapy, J. Instrum. 9 (2014) C12037, http://stacks.iop.org/1748-0221/9/i=12/a=C12037 
\bibitem{HIRFL_IonChamber}
Z. Xu, R. Mao, L. Duan, Q. She, Z. Hu, H. Li, Z. Lu, Q. Zhao, H. Yang, H. Su, C. Lu, R. Hu, J. Zhang, A new multi-strip ionization chamber used as online beam monitor for heavy ion therapy, Nucl. Instrum. Methods Phys. Res. Sec. A: Accelerators, Spectrometers, Detectors and Associated Equipment, 729, 21 November 2013, pp. 895-899, ISSN 0168-9002, http://dx.doi.org/10.1016/j.nima.2013.08.069.
\bibitem{INFN}
N. Givehchi, F. Marchetto, A. Boriano, A. Attili, F. Bourhaleb, R. Cirio, G.A.P. Cirrone, G. Cuttone, F. Di Rosa, M. Donetti, M.A. Garella, S. Giordanengo, S. Iliescu, A. La Rosa, P.A. Lojacono, P. Nicotra, C. Peroni, A. Pecka, G. Pitta, L. Raffaele, G. Russo, M.G. Sabini, L.M. Valastro, Online monitor detector for the protontherapy beam at the INFN Laboratori Nazionali del Sud-Catania, Nucl. Instrum. Methods Phys. Res. Sec. A: Accelerators, Spectrometers, Detectors and Associated Equipment, 572, 3, 21 March 2007, pp. 1094-1101, ISSN 0168-9002, http://dx.doi.org/10.1016/j.nima.2006.12.047.
\bibitem{CNAO}
S. Giordanengo, M. Donetti, M.A. Garella, F. Marchetto, G. Alampi, A. Ansarinejad, V. Monaco, M. Mucchi, I.A. Pecka, C. Peroni, R. Sacchi, M. Scalise, C. Tomba, R. Cirio, Design and characterization of the beam monitor detectors of the Italian National Center of Oncological Hadron-therapy (CNAO), Nucl. Instrum. Methods Phys. Res. Sec. A: Accelerators, Spectrometers, Detectors and Associated Equipment, 698, 11 January 2013, pp. 202-207, ISSN 0168-9002, http://dx.doi.org/10.1016/j.nima.2012.10.004.
\bibitem{TOP-IMPLART}
E. Basile, A. Carloni, D.M. Castelluccio, E. Cisbani, S. Colilli, G.De. Angelis, R. Fratoni, S. Frullani, F. Giuliani, M. Gricia, M. Lucentini, F. Santavenere and G. Vacca, An online proton beam monitor for cancer therapy based on ionization chambers with micro pattern readout,J. Instrum., 7(03), 2012, p. C03020, http://stacks.iop.org/1748-0221/7/i=03/a=C03020
\bibitem{maps}
R. Boll, M. Caccia, C.P. Welsch, M.H. Holzscheiter, Using Monolithic Active Pixel Sensors for fast monitoring of therapeutic hadron beams, Radiation Measurements, 46, 12, December 2011, pp. 1971-1973, ISSN 1350-4487, http://dx.doi.org/10.1016/j.radmeas.2011.05.053.
\bibitem{TPC}
M. Anderson, J. Berkovitz, W. Betts, R. Bossingham, F. Bieser, R. Brown, M. Burks, M. Calderón de la Barca Sánchez, D. Cebra, M. Cherney, J. Chrin, W.R. Edwards, V. Ghazikhanian, D. Greiner, M. Gilkes, D. Hardtke, G. Harper, E. Hjort, H. Huang, G. Igo, S. Jacobson, D. Keane, S.R. Klein, G. Koehler, L. Kotchenda, B. Lasiuk, A. Lebedev, J. Lin, M. Lisa, H.S. Matis, J. Nystrand, S. Panitkin, D. Reichold, F. Retiere, I. Sakrejda, K. Schweda, D. Shuman, R. Snellings, N. Stone, B. Stringfellow, J.H. Thomas, T. Trainor, S. Trentalange, R. Wells, C. Whitten, H. Wieman, E. Yamamoto, W. Zhang, The STAR time projection chamber: a unique tool for studying high multiplicity events at RHIC, Nucl. Instrum. Methods Phys. Res. Sec. A: Accelerators, Spectrometers, Detectors and Associated Equipment, Volume 499, Issues 2–3, 1 March 2003, pp. 659-678, ISSN 0168-9002, http://dx.doi.org/10.1016/S0168-9002(02)01964-2.
\bibitem{topmetal-II-}
M. An, C. Chen, C. Gao, M. Han, R. Ji, X. Li, Y. Mei, Q. Sun, X. Sun, K. Wang, L. Xiao, P. Yang, W. Zhou, A low-noise CMOS pixel direct charge sensor, Topmetal-II-, Nucl. Instrum. Methods Phys. Res. Sec. A: Accelerators, Spectrometers, Detectors and Associated Equipment, Volume 810, 21 February 2016, ISSN 0168-9002, pp. 144-150,  http://dx.doi.org/10.1016/j.nima.2015.11.153.
\bibitem{tm2-_lowtemp}
S. Zou, Y. Fan, M. An, C. Chen, G. Huang, J. Liu, H. Pei, X. Sun, P. Yang, D. Wang, L. Xiao, Z. Wang, K. Wang, W. Zhou, Test of Topmetal-II− In Liquid Nitrogen For Cryogenic Temperature TPCs, Nucl. Instrum. Methods Phys. Res. Sec. A: Accelerators, Spectrometers, Detectors and Associated Equipment, ISSN 0168-9002, http://dx.doi.org/10.1016/j.nima.2016.05.125.
\bibitem{negativeIon}
G.F. Knoll, Radiation Detection and Measurement, 3rd ed, John Wiley and sons Inc., New York, 2000, pp.131-132
\bibitem{fieldcage}
T. Behnke, K. Dehmelt, R. Diener, L. Steder, T. Matsuda, V. Prahl and P. Schade, A lightweight field cage for a large TPC prototype for the ILC, J. Instrum., 5(10), Pages p. 10011, http://stacks.iop.org/1748-0221/5/i=10/a=P10011
\bibitem{HIRFL_BeamDeliver}
Q. Li, Z. Dai, Z. Yan, X. Jin, X. Liu, G. Xiao, Heavy-ion conformal irradiation in the shallow-seated tumor therapy terminal at HIRFL, Med. Biol. Eng. Comput. 45 (11) (2007)  1037–1043, http://dx.doi.org/10.1007/s11517-007-0245-3
\bibitem{recombination}
T. Hiraoka, K. Kawashima and K. Hoshino, Ion recombination loss in ionization chambers irradiated by proton beams, Br. J. Radiol., 55(656)(1982) 585-587.
\end{thebibliography}
\end{document}